\documentclass[12pt]{article}
\usepackage{amssymb}
\date{}
\newcommand{\bee}{\begin{eqnarray}}
\newcommand{\ede}{\end{eqnarray}}
\begin{document}
\baselineskip=14 pt
\begin{center}
{\Large\bf Tortoise coordinate and Hawking effect in a dynamical
Kerr black hole }
\end{center}
\begin{center}
 Jian Yang${}^{a,}$\footnote{e-mail : yjccnuphy@yahoo.com.cn} Zheng Zhao
${}^{b,}$\footnote{e-mail : zhaoz43@hotmail.com}  Wenbiao Liu
${}^{b,}$\footnote{e-mail : wbliu@bnu.edu.cn}
\end{center}
\begin{center}
a, School of Science,\\
Beijing University of Posts and Telecommunications,\\
 Beijing, 100876, China, \\
 b, Department of Physics, Institute of theoretical physics,\\
Beijing Normal University,\\
 Beijing, 100875, China, \\

\end{center}
\begin{abstract}
 Hawking effect from a dynamical Kerr black hole is investigated using the improved Damour-Ruffini method with a new tortoise
 coordinate transformation. Hawking temperature of the black hole can be obtained point by
point at the event horizon. It is found that Hawking temperatures of
different points on the surface are different. Moreover, the
temperature does not turn to zero while the dynamical black hole
turns to an extreme one.
\end{abstract}



\section{Introduction}
It is well known that Hawking effect in a black hole is one of the
most striking phenomena\cite{1,2}. A black hole was recognized to
have thermal property right after the four laws of black hole
thermodynamics had been built successfully and Hawking radiation had
been discovered. In 1976, Damour and Ruffini proposed a new method
with which one can caculate Hawking radiation\cite{3}. Using this
method, Liu, et al have proved that a Kerr-Newman black hole
radiates Dirac particles\cite{4,5}. In 1990's, Z. Zhao, X. X. Dai
and Z. Q. Luo, et al improved Damour-Ruffini method to study Hawking
effect from some dynamical black holes. They only investigated
several kinds of dynamical spherically symmetric black holes via the
improved method\cite{6,7}.

Many works are also focused on Hawking effect from dynamical black
holes for these years~\cite{8,9}. Y. P. Zhang et al studied Hawking
effect from a Vaidya black hole via Hamilton-Jacobi
method~\cite{10}. S. W. Zhou et al discussed the same problem
through Parikh's tunneling method~\cite{11}. Recently we have
improved Zhao's tortoise coordinate. Using the new tortoise
coordinate, Hawking effect in some dynamical spherically symmetric
black holes has been investigated\cite{12}. We have got more
accurate expressions of surface gravity and Hawking temperature.

It is well known that a stationary black hole has identical
temperature on the event horizon surface. A dynamical spherically
symmetric black hole has also identical temperature at the same time
while its temperature varies with time\cite{12}. However, the
temperature at different points on the surface of an axisymmetric
dynamical black hole is probably a point-depending variable. Hawking
effect in a dynamical Kerr black hole will be investigated in the
following parts.

The organization is as follows. In Sec. 2, we will give a brief
overview on a dynamical Kerr black hole. In Sec. 3, we will discuss
Hawking effect in a dynamical Kerr black hole under the new tortoise
coordinate transformation. In Sec. 4, some conclusion and discussion
will be given.

\section{The dynamical Kerr black hole
}

The line element of a dynamical Kerr black hole can be expressed in
the advanced Eddington-Finkelstein time coordinate as \cite{13}
\begin{eqnarray}
& &
ds^{2}=-(1-\frac{2mr}{\rho^{2}})dv^{2}+2dvdr-\frac{4mra\sin^{2}\theta}{\rho^{2}}dvd\varphi
-2a\sin^{2}\theta drd\varphi \nonumber\\
& &
+\rho^{2}d\theta^{2}+(r^{2}+a^{2}+\frac{2mra^{2}\sin^{2}\theta}{\rho^{2}})\sin^{2}\theta
d\varphi^{2},
\end{eqnarray}
where $\rho^{2}=r^{2}+a^{2}\cos^{2}\theta,m=m(v),a=a(v)$.

The null hypersurface condition
\begin{equation}
g^{\mu\nu}\frac{\partial f}{\partial x^{\mu}}\frac{\partial
f}{\partial x^{\nu}}=0
\end{equation}
can be rewritten as
\begin{equation}
r^{2}(1-2\dot{r})-2mr+a^{2}(1-2\dot{r}+\dot{r}^{2}\sin^{2}\theta)+r'^{2}=0,\label{c1}
\end{equation}
where $\dot{r}=\frac{\partial r}{\partial v},r'=\frac{\partial
r}{\partial\theta}$.

The Eq.(\ref{c1}) determines the local event horizon of the
dynamical Kerr black hole, whose solution is
\begin{equation}
r_{H}=\frac{m\pm
\sqrt{m^{2}-(1-2\dot{r}_{H})[(1-2\dot{r}_{H}+\dot{r}_{H}^{2}\sin^{2}\theta)a^{2}
+r_{H}'^{2}]}}{1-2\dot{r}_{H}}.\label{c2}
\end{equation}
Obviously, it depends on the angular variable $\theta$, and is
different from the case of spherically symmetric black holes.
\section{Hawking effect from the event horizon}
The Klein-Gordon equation in a dynamical Kerr space-time is
\begin{eqnarray}
& & a^{2}\sin^{2}\theta\frac{\partial^{2}\Phi}{\partial
v^{2}}+2(a\dot{a}\sin^{2}\theta+r)\frac{\partial\Phi}{\partial
v}+2(r^{2}+a^{2})\frac{\partial^{2}\Phi}{\partial v\partial
r}\nonumber\\
& &  +2a\frac{\partial^{2}\Phi}{\partial v\partial
\varphi}+2a\frac{\partial^{2}\Phi}{\partial r\partial
\varphi}+(r^{2}+a^{2}-2mr)\frac{\partial^{2}\Phi}{\partial
r^{2}}\nonumber\\
& &  +2(r-m+a\dot{a})\frac{\partial\Phi}{\partial
r}+\frac{\partial^{2}\Phi}{\partial
\theta^{2}}+\dot{a}\frac{\partial\Phi}{\partial \varphi}\nonumber\\
& & +\cot\theta\frac{\partial\Phi}{\partial
\theta}+\frac{1}{\sin^{2}\theta}\frac{\partial^{2}\Phi}{\partial
\varphi^{2}}-\mu^{2}\rho^{2}\Phi=0,\label{c3}
\end{eqnarray}
where $\mu$ is the mass of a Klein-Gordon particle.

It is well known that when Hawking effect from a Schwarzschild black
hole is investigated using Damour-Ruffini method, the tortoise
coordinate is defined as following\cite{3}
\begin{equation}
r_{\ast}=r+2M\ln[\frac{r-2M}{2M}].
\end{equation}
For a dynamical Kerr black hole, a new tortoise coordinate can be
written as
\begin{eqnarray}
& &
r_{\ast}=r+\frac{1}{2\kappa(v_{0},\theta_{0})}\ln[\frac{r-r_{H}(v,\theta)}{r_{H}(v,\theta)}],
\nonumber\\
& &v_{\ast}=v-v_{0}, \theta_{\ast}=\theta-\theta_{0},
\end{eqnarray}
where both $v_{0}$ and $\theta_{0}$ are constants under tortoise
coordinate transformation. At the same time, $v_{0}$ is the moment
when the particle escapes from the event horizon of the black hole
and depicts the evolution of the black hole, $\theta_{0}$ is the
location where the particle escapes from the event horizon of the
black hole and depicts the shape of the black hole. According to the
tortoise coordinate transformation, we have
\begin{eqnarray*}
& &  \frac{\partial}{\partial
r}=[1+\frac{1}{2\kappa(r-r_{H})}]\frac{\partial}{\partial
r_{\ast}},\\
& &  \frac{\partial}{\partial v}=\frac{\partial}{\partial
v_{\ast}}-\frac{r\dot{r}_{H}}{2\kappa
r_{H}(r-r_{H})}\frac{\partial}{\partial
r_{\ast}},\\
& &  \frac{\partial}{\partial \theta}=\frac{\partial}{\partial
\theta_{\ast}}-\frac{rr_{H}'}{2\kappa
r_{H}(r-r_{H})}\frac{\partial}{\partial r_{\ast}},
\end{eqnarray*}
\begin{eqnarray*}
& & \frac{\partial^{2}}{\partial
r^{2}}=[1+\frac{1}{2\kappa(r-r_{H})}]^{2}\frac{\partial^{2}}{\partial
r_{\ast}^{2}}-\frac{1}{2\kappa(r-r_{H})^{2}}\frac{\partial}{\partial
r_{\ast}},\\
& & \frac{\partial^{2}}{\partial r \partial
v}=[1+\frac{1}{2\kappa(r-r_{H})}]\frac{\partial^{2}}{\partial
r_{\ast}
\partial v_{\ast}}+\frac{\dot{r}_{H}}{2\kappa(r-r_{H})^{2}}\frac{\partial}{\partial
r_{\ast}}\\
& & -\frac{r\dot{r}_{H}}{2\kappa
r_{H}(r-r_{H})}[1+\frac{1}{2\kappa(r-r_{H})}]
\frac{\partial^{2}}{\partial r_{\ast}^{2}},
\end{eqnarray*}
\begin{eqnarray*}
& & \frac{\partial^{2}}{\partial
v\partial\varphi}=\frac{\partial^{2}}{\partial
v_{\ast}\partial\varphi}-\frac{r\dot{r}_{H}}{2\kappa
r_{H}(r-r_{H})}\frac{\partial^{2}}{\partial
r_{\ast}\partial\varphi},\\
& &  \frac{\partial^{2}}{\partial
r\partial\varphi}=[1+\frac{1}{2\kappa(r-r_{H})}]\frac{\partial^{2}}{\partial
r_{\ast}\partial\varphi},
\end{eqnarray*}
\begin{eqnarray*}
& & \frac{\partial^{2}}{\partial v^{2}}=\frac{\partial^{2}}{\partial
v_{\ast}^{2}}-\frac{r\ddot{r}_{H}r_{H}(r-r_{H})-r\dot{r}_{H}^{2}(r-r_{H})+r\dot{r}_{H}^{2}r_{H}}
{2\kappa r_{H}^{2}(r-r_{H})^{2}}\frac{\partial}{\partial r_{\ast}}
\\
& & -\frac{2r\dot{r}_{H}}{2\kappa
r_{H}(r-r_{H})}\frac{\partial^{2}}{\partial r_{\ast}\partial
v_{\ast}}+\frac{r^{2}\dot{r}_{H}^{2}}{4\kappa^{2}
r_{H}^{2}(r-r_{H})^{2}}\frac{\partial^{2}}{\partial
r_{\ast}^{2}},\\
& & \frac{\partial^{2}}{\partial
\theta^{2}}=\frac{\partial^{2}}{\partial
\theta_{\ast}^{2}}-\frac{rr_{H}''r_{H}(r-r_{H})-rr_{H}'^{2}(r-r_{H})+rr_{H}'^{2}r_{H}}
{2\kappa r_{H}^{2}(r-r_{H})^{2}}\frac{\partial}{\partial r_{\ast}}
\\
& & -\frac{2rr_{H}'}{2\kappa
r_{H}(r-r_{H})}\frac{\partial^{2}}{\partial r_{\ast}\partial
\theta_{\ast}}+\frac{r^{2}r_{H}'^{2}}{4\kappa^{2}
r_{H}^{2}(r-r_{H})^{2}}\frac{\partial^{2}}{\partial r_{\ast}^{2}}.
\end{eqnarray*}
The Klein-Gordon Eq.(\ref{c3}) can be rewritten as
\begin{eqnarray}
& &
\frac{a^{2}r^{2}\dot{r}_{H}^{2}\sin^{2}\theta-2r_{H}(r^{2}+a^{2})[1+2\kappa(r-r_{H})]r\dot{r}_{H}}{
2\kappa
r_{H}(r-r_{H})\{r_{H}(r^{2}+a^{2})[1+2\kappa(r-r_{H})]-a^{2}r\dot{r}_{H}\sin^{2}\theta\}}
\frac{\partial^{2}\Phi}{\partial r_{\ast}^{2}}\nonumber\\
& &
+\frac{(r^{2}+a^{2}-2mr)[1+2\kappa(r-r_{H})]^{2}r_{H}^{2}+r^{2}r_{H}'^{2}}{
2\kappa
r_{H}(r-r_{H})\{r_{H}(r^{2}+a^{2})[1+2\kappa(r-r_{H})]-a^{2}r\dot{r}_{H}\sin^{2}\theta\}}
\frac{\partial^{2}\Phi}{\partial r_{\ast}^{2}}\nonumber\\
& & +2\frac{\partial^{2}\Phi}{\partial v_{\ast}\partial
r_{\ast}}+\frac{1}{r_{H}(r^{2}+a^{2})[1+2\kappa(r-r_{H})]-a^{2}r\dot{r}_{H}\sin^{2}\theta}\{-a^{2}\sin^{2}\theta\nonumber\\
& &
\times\frac{r\ddot{r}_{H}r_{H}(r-r_{H})-r(r-r_{H})\dot{r}_{H}^{2}+r\dot{r}_{H}^{2}r_{H}}{r_{H}(r-r_{H})}
-2r\dot{r}_{H}(a\dot{a}\sin^{2}\theta+r)\nonumber\\
& &
+\frac{2(r^{2}+a^{2})\dot{r}_{H}r_{H}}{r-r_{H}}-\frac{r_{H}(r^{2}+a^{2}-2mr)}{r-r_{H}}+2r_{H}(r-m+a\dot{a})
[1+2\kappa(r-r_{H})]\nonumber\\
& &
-\frac{rr_{H}r_{H}''(r-r_{H})-rr_{H}'^{2}(r-r_{H})+rr_{H}'^{2}r_{H}}{r_{H}(r-r_{H})}
-rr_{H}'\cot\theta\}\frac{\partial\Phi}{\partial
r_{\ast}}\nonumber\\
& & +\frac{2ar_{H}[1+2\kappa(r-r_{H})]-2ar\dot{r}_{H}}
{r_{H}(r^{2}+a^{2})[1+2\kappa(r-r_{H})]-a^{2}r\dot{r}_{H}\sin^{2}\theta}
\frac{\partial^{2}\Phi}{\partial
r_{\ast}\partial\varphi}\nonumber\\
& &
-\frac{2rr_{H}'}{r_{H}(r^{2}+a^{2})[1+2\kappa(r-r_{H})]-a^{2}r\dot{r}_{H}
\sin^{2}\theta}\frac{\partial^{2}\Phi}{\partial
r_{\ast}\partial\theta_{\ast}}\nonumber\\
& &
+\frac{2\kappa(r-r_{H})r_{H}}{r_{H}(r^{2}+a^{2})[1+2\kappa(r-r_{H})]-a^{2}r\dot{r}_{H}\sin^{2}\theta}[a^{2}\sin^{2}\theta
\frac{\partial^{2}\Phi}{\partial
v_{\ast}^{2}}\nonumber\\
& & +2(a\dot{a}\sin^{2}\theta+r)\frac{\partial\Phi}{\partial
v_{\ast}}+\frac{\partial^{2}\Phi}{\partial
\theta_{\ast}^{2}}+\dot{a}\frac{\partial\Phi}{\partial\varphi}+\cot\theta\frac{\partial\Phi}{\partial
\theta_{\ast}}\nonumber\\
& & +2a\frac{\partial^{2}\Phi}{\partial
v_{\ast}\partial\varphi}+\frac{1}{\sin^{2}\theta}\frac{\partial^{2}\Phi}{\partial\varphi^{2}}-\mu^{2}\rho^{2}\Phi]=0.\label{c4}
\end{eqnarray}
From the null hypersurface condition Eq.(\ref{c1}) of a dynamical
Kerr space-time, the numerator of the coefficient of the term
$\frac{\Phi^{2}\rho}{\partial r_{\ast}^{2}}$ approaches to zero at
the event horizon $r_{H}$. Therefore we can calculate the limit of
the coefficient using L'Hospital law. Let the limit to be equal to
an undetermined constant $K$
\begin{eqnarray*}
& & \lim_{r\rightarrow r_{H},v\rightarrow
v_{0},\theta\rightarrow\theta_{0}}\frac{a^{2}r^{2}\dot{r}_{H}^{2}\sin^{2}\theta-2r_{H}(r^{2}+a^{2})[1+2\kappa(r-r_{H})]r\dot{r}_{H}}{
2\kappa
r_{H}(r-r_{H})\{r_{H}(r^{2}+a^{2})[1+2\kappa(r-r_{H})]-a^{2}r\dot{r}_{H}\sin^{2}\theta\}}\\
& & +\lim_{r\rightarrow r_{H},v\rightarrow
v_{0},\theta\rightarrow\theta_{0}}\frac{(r^{2}+a^{2}-2mr)[1+2\kappa(r-r_{H})]^{2}r_{H}^{2}+r^{2}r_{H}'^{2}}{
2\kappa
r_{H}(r-r_{H})\{r_{H}(r^{2}+a^{2})[1+2\kappa(r-r_{H})]-a^{2}r\dot{r}_{H}\sin^{2}\theta\}}\\
& &    =K,
\end{eqnarray*}
while $\kappa$ is selected as
\begin{eqnarray}
& &
\kappa=\frac{(1-2\dot{r}_{H})r_{H}-m}{4mr_{H}-(1-2\dot{r}_{H})(r_{H}^{2}+a^{2})-\dot{r}_{H}a^{2}\sin^{2}\theta_{0}}\nonumber\\
& &
+\frac{\frac{\dot{r}_{H}^{2}a^{2}\sin^{2}\theta_{0}-(r_{H}^{2}+a^{2})\dot{r}_{H}}{r_{H}}}{4mr_{H}-(1-2\dot{r}_{H})(r_{H}^{2}+a^{2})-\dot{r}_{H}a^{2}\sin^{2}\theta_{0}},
\end{eqnarray}
we have $K=1$.

From Eq.(\ref{c2}), external horizon $r_{+}$ is given by
\begin{displaymath}
r_{+}=\frac{m+
\sqrt{m^{2}-(1-2\dot{r}_{+})[(1-2\dot{r}_{+}+\dot{r}_{+}^{2}\sin^{2}\theta_{0})a^{2}
+r_{+}'^{2}]}}{1-2\dot{r}_{+}}.
\end{displaymath}
When $r$ approaches to $r_{+}$, the Klein-Gordon Eq.(\ref{c4}) can
be transformed into
\begin{equation}
\frac{\partial^{2}\Phi}{\partial
r_{\ast}^{2}}+2\frac{\partial^{2}\Phi}{\partial v_{\ast}\partial
r_{\ast}}+A\frac{\partial\Phi}{\partial
r_{\ast}}+B\frac{\partial^{2}\Phi}{\partial
r_{\ast}\partial\varphi}+C\frac{\partial^{2}\Phi}{\partial
r_{\ast}\partial\theta_{\ast}}=0,\label{c5}
\end{equation}
where
\begin{eqnarray}
A&=&\frac{2r_{+}\dot{r}_{+}-\ddot{r}_{+}a^{2}\sin^{2}\theta_{0}+2a\dot{a}(1-\dot{r}_{+}\sin^{2}\theta_{0})
-r_{+}''-r_{+}'\cot\theta_{0}}{r_{+}^{2}+a^{2}(1-\dot{r}_{+}\sin^{2}\theta_{0})},\nonumber\\
B&=&\frac{2a(1-\dot{r}_{+})}{r_{+}^{2}+a^{2}(1-\dot{r}_{+}\sin^{2}\theta_{0})},\nonumber\\
C&=&\frac{-2r_{+}'}{r_{+}^{2}+a^{2}(1-\dot{r}_{+}\sin^{2}\theta_{0})}.
\end{eqnarray}
Separate variables as following
\begin{equation}
\Phi=R(r_{\ast})\Theta(\theta_{\ast})e^{il\varphi-i\omega v_{\ast}},
\end{equation}
where $\omega$ is the energy of Klein-Gordon particle, $l$ is the
projection of angular momentum on $\varphi$-axis.We can get
\begin{eqnarray}
& & \Theta'=\lambda\Theta,\nonumber\\
& & R''+(A+\lambda C+ilB-2i\omega)R'=0,\label{c6}
\end{eqnarray}
where the constant $\lambda$ is introduced by the separation of
variables.

Assuming
$\lambda=\lambda_{1}+i\lambda_{2},\lambda_{1},\lambda_{2}\in R$, the
Eq.(\ref{c6}) can be rewritten as
\begin{eqnarray}
& & \Theta'=(\lambda_{1}+i\lambda_{2})\Theta,\nonumber\\
& & R''+[A+( \lambda_{1}+i\lambda_{2})C+ilB-2i\omega]R'=0,
\end{eqnarray}
and its solution is
\begin{eqnarray}
& & \Theta=c_{1}e^{(\lambda_{1}+i\lambda_{2})\theta_{\ast}},\nonumber\\
& & R=c_{2}e^{-[A+(
\lambda_{1}+i\lambda_{2})C+ilB-2i\omega]r_{\ast}}+c_{3},
\end{eqnarray}
where $c_{1},c_{2}$ and $c_{3}$ are integral constants,
$\theta_{\ast}$ is polar angle. Its radial ingoing and outgoing
components are respectively
\begin{eqnarray}
\psi_{in}&=&e^{-i\omega v_{\ast}},\nonumber\\
\psi_{out}&=&e^{-i\omega
v_{\ast}}e^{2i(\omega-\omega_{0})r_{\ast}}e^{-(A+\lambda_{1}
C)r_{\ast}},
\end{eqnarray}
where
\begin{equation}
\omega_{0}=\frac{1}{2}lB-\frac{1}{2}\lambda_{2}C=\frac{la(1-\dot{r}_{+})+\lambda_{2}
r_{+}'}{r_{+}^{2}+a^{2}(1-\dot{r}_{+}\sin^{2}\theta_{0})}.
\end{equation}
The outgoing wave can be rewritten as
\begin{equation}
\psi_{out}=e^{-i\omega
v_{\ast}}e^{2i(\omega-\omega_{0})r}e^{-\bar{A}r}(\frac{r-r_{+}}{r_{+}})^{\frac{i(\omega-\omega_{0})}
{\kappa}}(\frac{r-r_{+}}{r_{+}})^{-\frac{\bar{A}}{2\kappa}},
\end{equation}
where $\bar{A}=A+\lambda_{1} C$. It is obvious that the outgoing
wave is not analytical at the horizon $r_{+}$. Extending the
outgoing wave from outside to inside of the horizon analytically
through the negative half complex plane, we get
\begin{equation}
\tilde{\psi}_{out}=e^{-i\omega
v_{\ast}}e^{2i(\omega-\omega_{0})r_{\ast}}e^{-\bar{{A}}r_{\ast}}e^{\frac{i\pi\bar{A}}{2\kappa}}
e^{\frac{\pi(\omega-\omega_{0})}{\kappa}}.
\end{equation}
The scattering probability of outgoing wave at the horizon is
\begin{equation}
\left\vert\frac{\psi_{out}}{\tilde{\psi}_{out}}\right\vert^{2}=e^{-\frac{2\pi(\omega-\omega_{0})}{\kappa}}.
\end{equation}
According to the explanation of Sannan\cite{14}, the outgoing wave
has black body spectrum
\begin{eqnarray}
N_{\omega}&=&\frac{1}{e^{\frac{\omega-\omega_{0}}{k_{B}T}}\pm1},\\
T&=&\frac{\kappa}{2\pi k_{B }}\label{c7}.
\end{eqnarray}
\section{Conclusion and Discussion}
The Hawking effect from a dynamical Kerr black hole has been studied
under a new tortoise coordinate transformation. It is found that
Hawking temperature $T$ not only depends on time but also varies
with angle. Because there is temperature gradient between equator
and pole, the heat fluid should exist on the black hole event
horizon surface. It is an interesting issue which deserves to be
studied further.

According to expression of local event horizon Eq.(\ref{c2}), the
external horizon $r_{+}$ and the internal horizon $r_{-}$ are given
respectively:
\begin{eqnarray}
& & r_{+}=\frac{m+
\sqrt{m^{2}-(1-2\dot{r}_{+})[(1-2\dot{r}_{+}+\dot{r}_{+}^{2}\sin^{2}\theta_{0})a^{2}
+r_{+}'^{2}]}}{1-2\dot{r}_{+}}, \nonumber\\
& & r_{-}=\frac{m-
\sqrt{m^{2}-(1-2\dot{r}_{-})[(1-2\dot{r}_{-}+\dot{r}_{-}^{2}\sin^{2}\theta_{0})a^{2}
+r_{-}'^{2}]}}{1-2\dot{r}_{-}}.
\end{eqnarray}
When the external horizon coincides with the internal one, i.e.
$r_{+}=r_{-}=r_{H}$, we will have
\begin{eqnarray*}
& & (1-2\dot{r}_{H})r_{H}-m=0, \\
& &
m^{2}-(1-2\dot{r}_{H})[(1-2\dot{r}_{H}+\dot{r}_{H}^{2}\sin^{2}\theta)a^{2}
+r_{H}'^{2}]=0.
\end{eqnarray*}
Now the surface gravity is equal to
\begin{eqnarray}
  \kappa
&=&\frac{\frac{\dot{r}_{H}^{2}a^{2}\sin^{2}\theta_{0}-(r_{H}^{2}+a^{2})\dot{r}_{H}}{r_{H}}}{4mr_{H}-(1-2\dot{r}_{H})(r_{H}^{2}+a^{2})-\dot{r}_{H}a^{2}\sin^{2}\theta_{0}}
\nonumber\\
&=&\frac{\frac{(1-2\dot{r}_{H})(r_{H}^{2}-a^{2})-r_{H}'^{2}-(r_{H}^{2}+a^{2})\dot{r}_{H}}{r_{H}}}{4mr_{H}-(1-2\dot{r}_{H})(r_{H}^{2}+a^{2})-\dot{r}_{H}a^{2}\sin^{2}\theta_{0}},
\end{eqnarray}
and this is the temperature of the extreme dynamical black hole.
When $\dot{r}_{H}=0$, the temperature will be equal to zero, which
is consistent with the case of a stationary Kerr black hole.

\section*{Acknowledgement}
One of the authors, Jian Yang, would like to thank Dr. Shiwei Zhou
and Dr. Xianming Liu for their helpful discussions . This research
is supported by the National Natural Science Foundation of
China(Grant Nos. 10773002,10875012,10875018).

\end{document}